\documentclass{iau} 
\usepackage{graphicx}

\newcommand{\hii}    {H\,{\sc{ii}}~}

\title[Infrared appearance of windblown bubbles] %% give here short title %%
{Infrared appearance of wind-blown bubbles around young massive stars}

\author[Maria S. Kirsanova \& Yaroslav N. Pavlyuchenkov]   %% give here short author list %%
{Maria S. Kirsanova$^1$
 \and Yaroslav N. Pavlyuchenkov$^1$}

\affiliation{$^1$Institute of Astronomy, Russian Academy of Sciences, \\ 119017, 48 Pyatnitskaya str., Moscow, Russia \\ email: {\tt kirsanova@inasan.ru} }

\pubyear{2022}
\volume{xxx}  %% insert here IAU Symposium No.
\jname{Predictive Power of Computational Astrophysics as a Discovery Tool}
\editors{A.C. Editor, B.D. Editor \& C.E. Editor, eds.}
\begin{document}

\maketitle

\begin{abstract}

Thousands of ring-like bubbles appear on infrared images of the Galaxy plane. Most of these infrared bubbles form during expansion of \hii regions around massive stars. However, the physical effects that determine their morphology are still under debate. Namely, the absence of the infrared emission toward the centres of the bubbles can be explained by pushing the dust grains by stellar radiation pressure. At the same time, small graphite grains and PAHs are not strongly affected by the radiation pressure and must be removed by another process. Stellar ultraviolet emission can destroy the smallest PAHs but the photodestruction is ineffective for the large PAHs. Meanwhile, the stellar wind can evacuate all types of grains from \hii{} regions. In the frame of our chemo-dynamical model we vary parameters of the stellar wind and illustrate their influence on the morphology and synthetic infrared images of the bubbles.

\end{abstract}

\firstsection % if your document starts with a section,
\section{Introduction}

\hii{} regions around young OB-type stars represent natural laboratories to study properties of cosmic dust particles. Observations made with {\it Spitzer} and {\it WISE} telescope revealed different spatial morphology of the middle and far-infrared (IR) emission in \hii regions, see e.g.~\cite{Churchwell_etal06, Deharveng_etal10, Anderson_etal14}. The mid-IR emission of these objects looks like a compact arc or a small ring at $24\,\mu$m, surrounded by a larger ring of emission at $8\,\mu$m, see e.g.~\cite{Simpson_etal12}. The larger ring is observed toward the dense envelope of neutral gas, swept up by a shock front from an expanding \hii{} region. Therefore, these objects are often called as infrared bubbles. Subsequent observations with {\em Herschel\/} have demonstrated that the ring-like appearance is also typical for far-IR emission, see e.g.~\cite{Molinari_etal10, Anderson_etal12}. Series of papers by \cite{Pavyar_etal13, Akimkin_etal15, Akimkin_etal17} aims to study different factors which determine quantitatively and qualitatively spatial distribution and intensity of the mid and far-IR emission from the \hii{} regions and surrounding neutral material. They found that none of the following factors: photo-destruction by ultraviolet photons, radiation pressure and overall expansion of \hii{} regions can explain all the observed IR features alone. Recent observations of \hii{} regions by {\it SOFIA} telescope revealed the importance of stellar wind in dynamics of ionised gas around O-type stars, see e.~g. \cite{Pabst_etal19, Luisi_etal21}. Following these results, we numerically explore the influence of the wind on dust dynamics and simulate mid and far-IR emission from \hii{} regions. Our main aim is to find out if the inclusion of wind allows us to reproduce all the observed IR features simultaneously. Moreover, except the \hii{} regions, our calculations can be useful to interpret mid-IR observations of old stellar populations: such as luminous blue variables or Wolf-Rayet stars, see e.g.~\cite{Gvaramadze_etal10, Wachter_etal10}.

\section{Simulations}

We considered three models with different parameters of stellar wind in order to demonstrate its crucial role in dynamics of dust grains. The three models have common parameters such as spectral type of ionising star, O7\,V with effective temperature 37000~K, and initial uniformly-distributed gas number density $n_0=10^3$~cm$^{-3}$. The dust-to-gas mass ratio has standard value 1/100, and the dust is uniformly mixed with gas in the beginning of the simulations. First model has no stellar wind. It is needed to compare results of two other models with different parameters of the wind. For this model, we estimated thermal energy $E_{\rm therm}$ of the HII region at particular moment of model time when radius of the ionized bubble reached 0.7~pc. In the second model the wind is ''weak'', where energy that brings stellar wind is comparable with $E_{\rm therm}$, i. e. $E_{\rm wind} \approx E_{\rm therm}$. The wind mass-loss rate in this model is to be $dM/dt = 4\times 10^{-8}$~M$\odot$~yr$^{-1}$ and the wind terminal velocity $v_{\infty} = 500$~km~s$^{-1}$. The third model has ''strong wind'', where $E_{\rm wind} \approx 100 E_{\rm therm}$, $dM/dt = 1.5\times 10^{-7}$~M$\odot$~yr$^{-1}$ and $v_{\infty} = 2000$~km~s$^{-1}$. To calculate the radiation intensity distributions we adopt the same radiative transfer model as in \cite{Pavyar_etal13}. This model takes into account the stochastic (transient) heating of PAHs and small grains by single photons. The thermal state of a grain of particular type is described by the probability density distribution over the temperature, $P(T)$, which represents a fraction of the grains having temperature in the range $(T,T+dT)$.

\section{Results}

\begin{figure}
    \begin{center}
    \includegraphics[width=0.49\columnwidth]{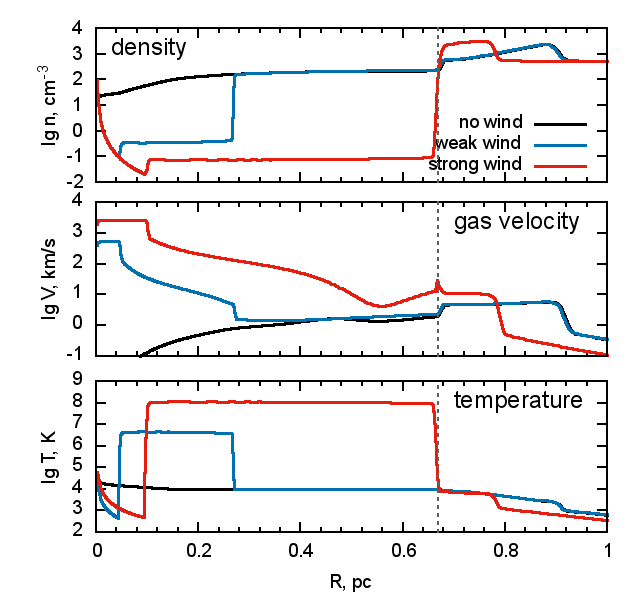}
    \includegraphics[width=0.49\columnwidth]{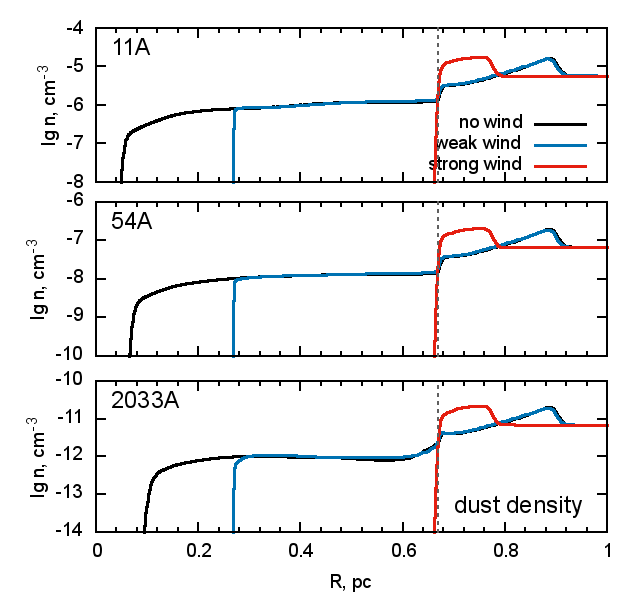}
    \caption{Summary of the model physical structure of the HII region at the time, when its radius is $\approx 0.7$~pc. Left: distributions of gas density (top), velocity (middle), and gas temperature (bottom). Right: distributions of dust number densities for the  considered  dust  components:  PAH  (11\AA, top),  graphite (54\AA, middle),  and  silicate grain (0.2$\mu$m~bottom).}
    \label{figphys}
    \end{center}
\end{figure}

A general physical structure of the HII region and surrounding gas at the moment when its radius is $\approx 0.7$~pc is shown in Fig.~\ref{figphys}.  An ionised gas region is surrounded by a dense envelope of neutral hydrogen (atomic and molecular), which  has  been  shovelled by the shock preceding the ionization front. The ionised region is rarefied and appears as a hot bubble in the strong wind model, while the empty cavity occupies less than half of the ionised volume in the weak wind model. Stellar wind makes expansion of the ionised region faster, therefore the age of the HII region is 59, 58 and 31 thousand~yr in the models without wind, with the weak and strong wind, respectively. The gas velocity is up to order of magnitude higher (10 vs 1~km~s$^{-1}$) in the strong wind model at the ionisation front and in the dense neutral envelope. Radial distributions of dust grains with different sizes have different radii of the empty inner cavity in the model without wind, namely the larger the grain the larger the cavity. PAH particles, having the largest surface-to-mass ratio, are well coupled to the gas in contrast with the big grains, as we discussed in \cite{Akimkin_etal15}. There is no segregation of the grain ensemble in the models with the wind, and the radii of the inner cavities are determined by the impetus transfer from the wind to the gas and dust medium.

\begin{figure}
    \begin{center}
    \includegraphics[width=0.32\columnwidth]{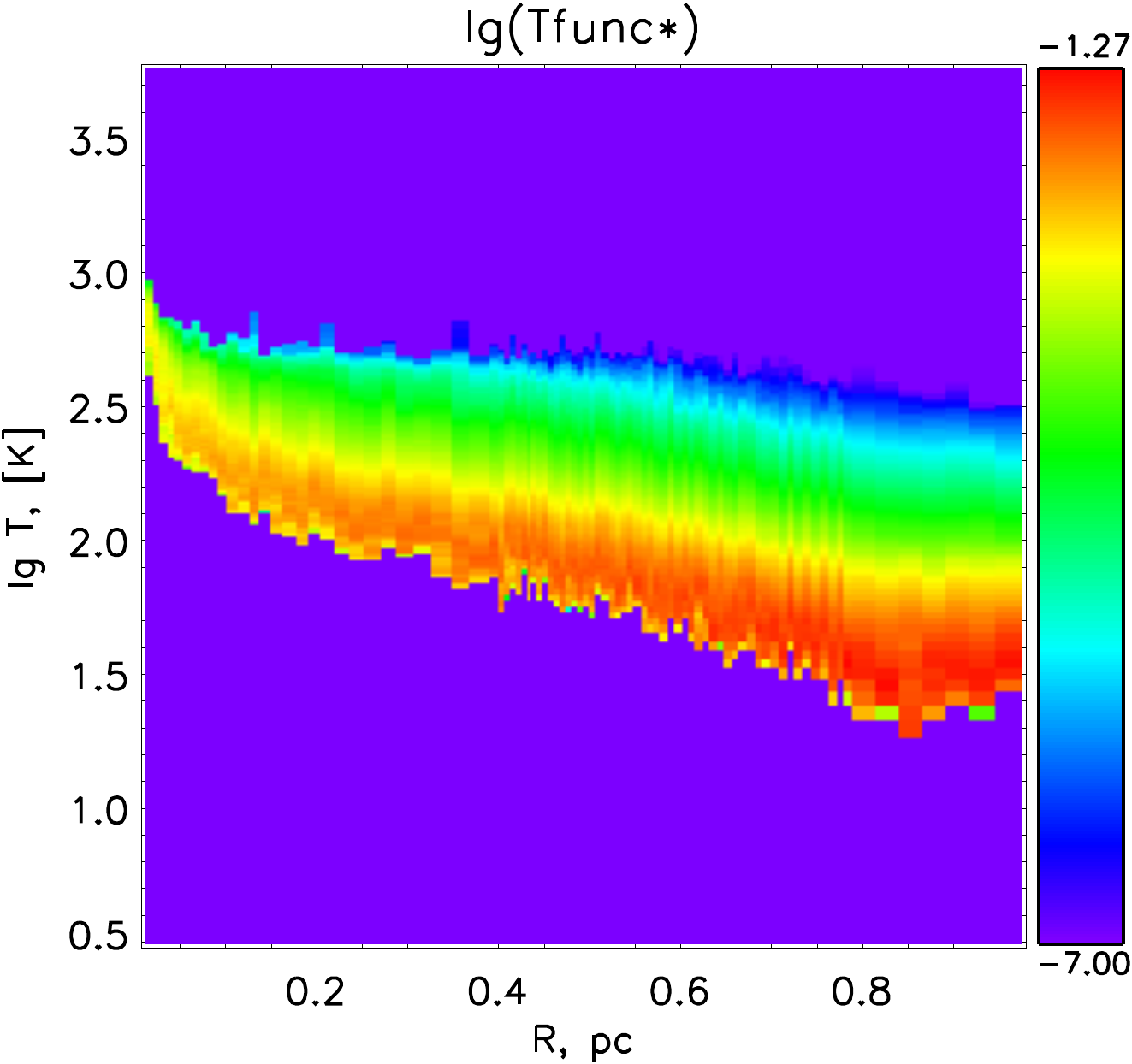}
    \includegraphics[width=0.32\columnwidth]{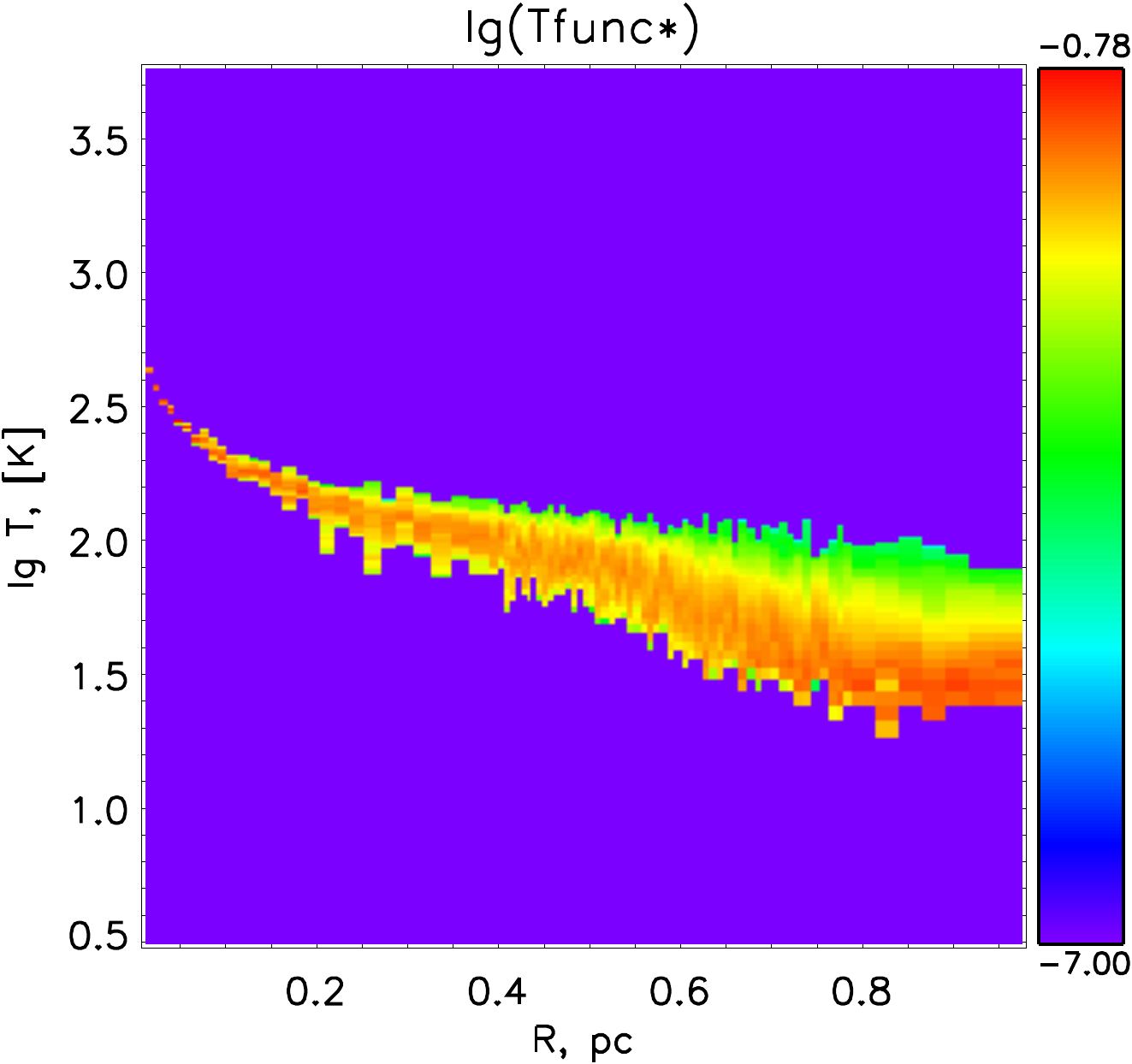}
    \includegraphics[width=0.32\columnwidth]{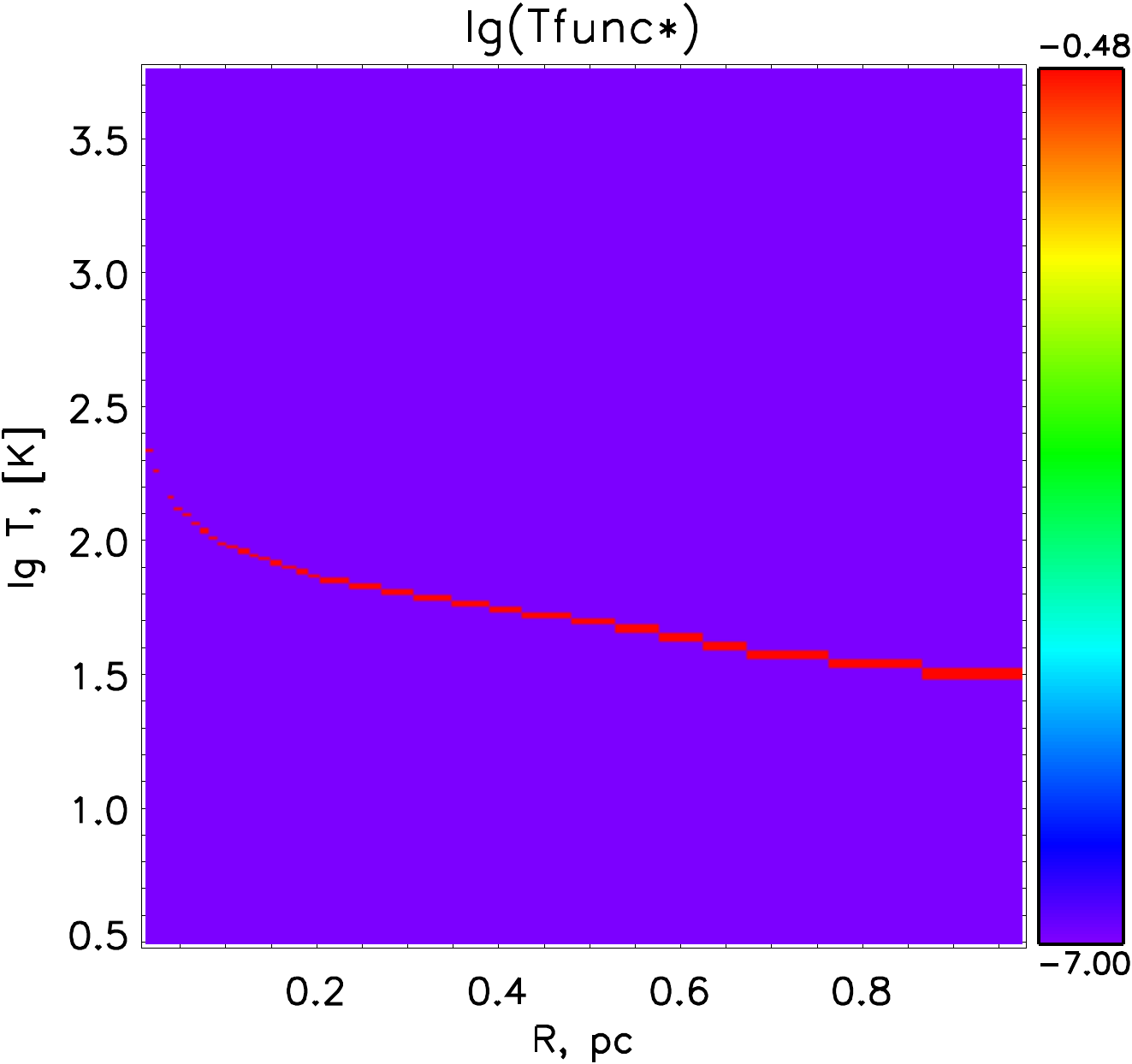}
    \caption{Radial dependence of temperature probability, $P(T)$, for the selected dust grain types: PAHs (11\AA), small graphites (middle, 50\AA),and big graphites (right, 0.2$\mu$m).}
    \label{figheat}
    \end{center}
\end{figure}

We illustrate the importance of stochastic heating in Fig.~\ref{figheat} which shows how $P(T)$ depends on the distance from the star for selected dust components which correspond to PAHs, very small graphite grains (VSGs),  and big graphite grains (BGs). The distribution of $P(T)$ for PAHs is broad over the entire cloud. In the very centre of the \hii region, the VSGs are heated up to 1000~K, and the temperature distribution is quite narrow, because under the strong radiation from the ionising star these grains tend to have equilibrium temperatures despite their relatively small size. Further from the star, the temperature distribution becomes wider, so that in the middle of the ionised region the temperature of VSGs fluctuates approximately from 50\,K to 200\,K. The $P(T)$ for big grains is represented by delta-function everywhere in the considered region, so the BGs temperature can be reliably described in the framework of the thermal equilibrium.

\begin{figure}
    \begin{center}
    \includegraphics[width=0.45\columnwidth]{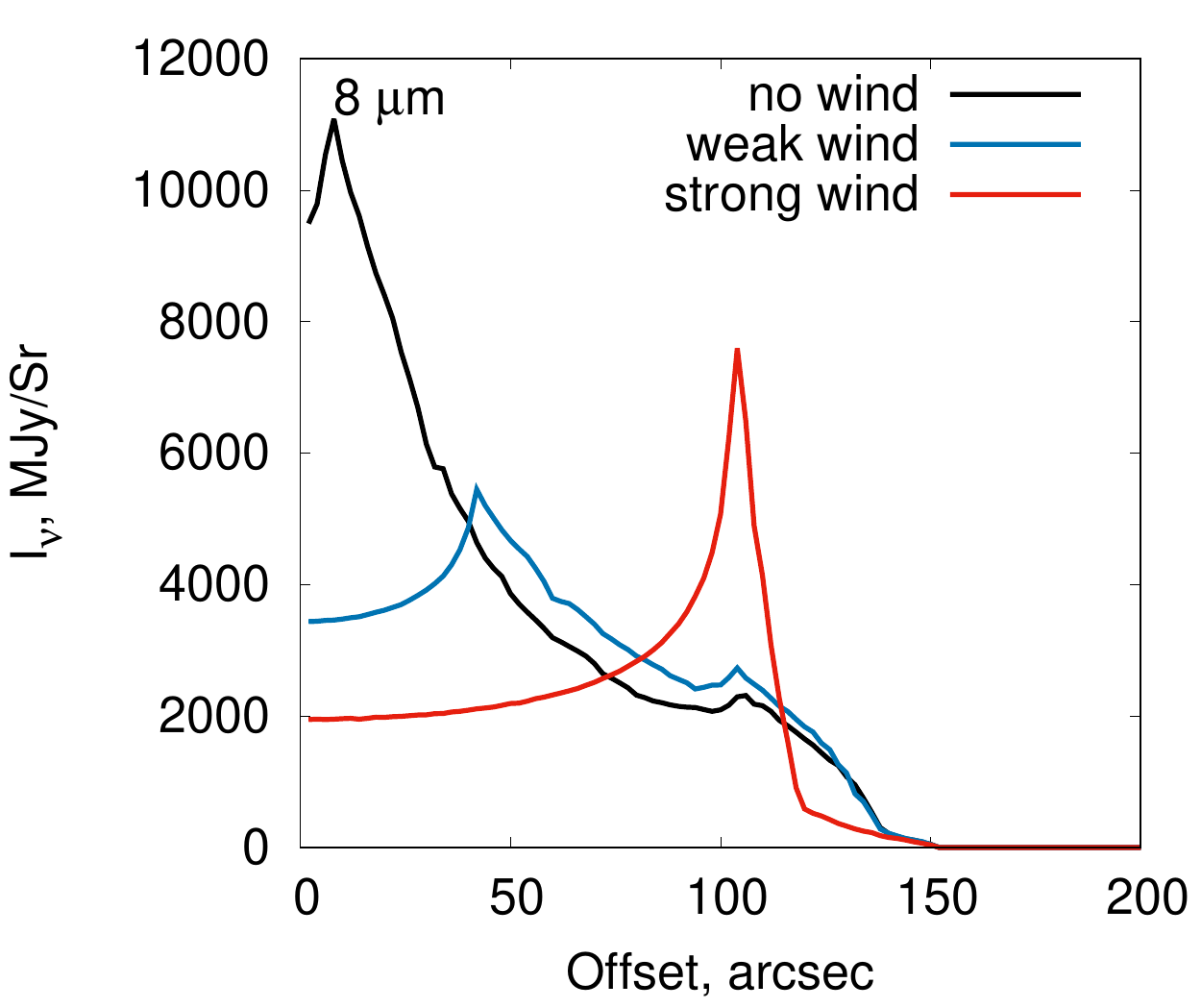}
    \includegraphics[width=0.45\columnwidth]{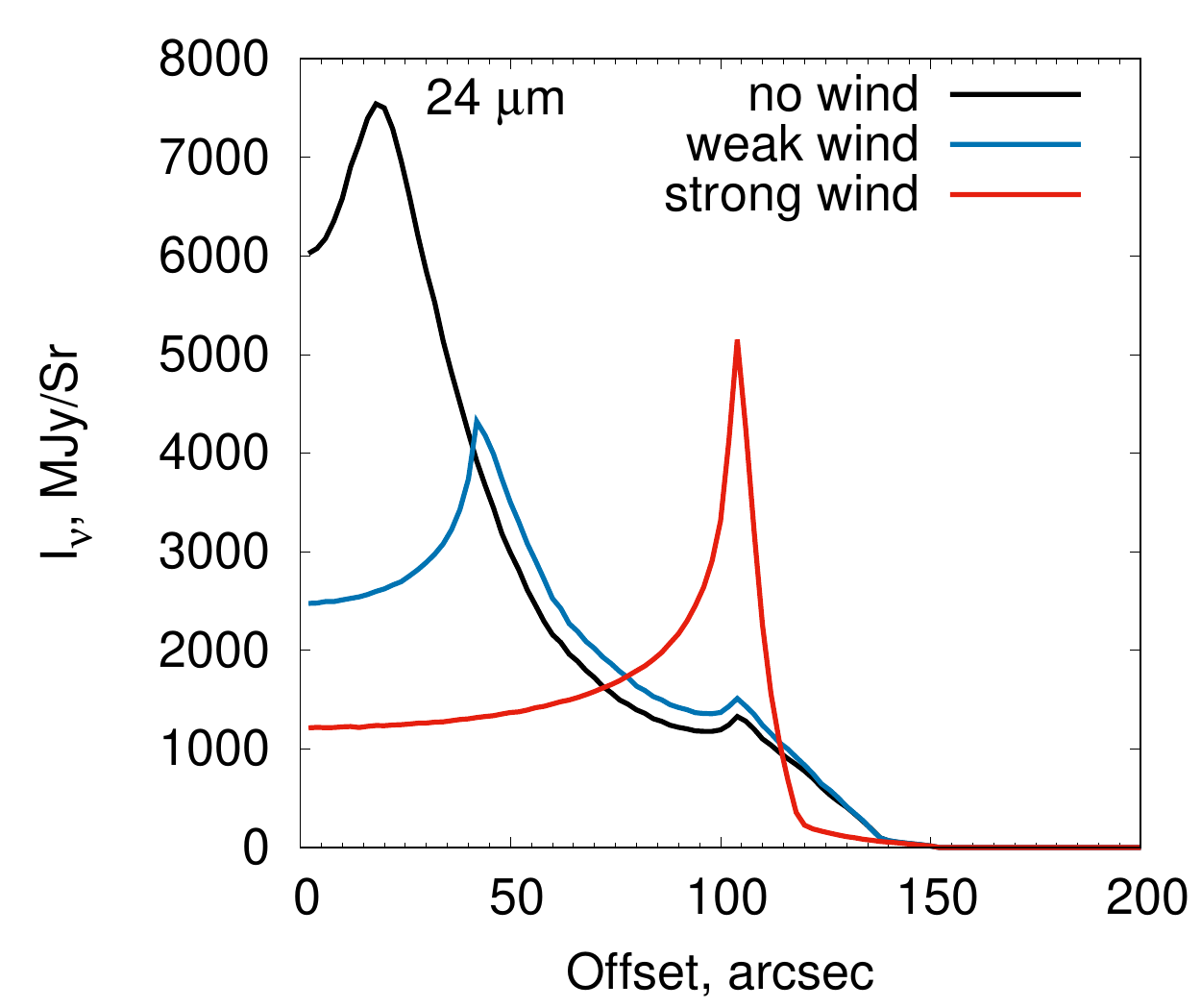}\\
    \includegraphics[width=0.45\columnwidth]{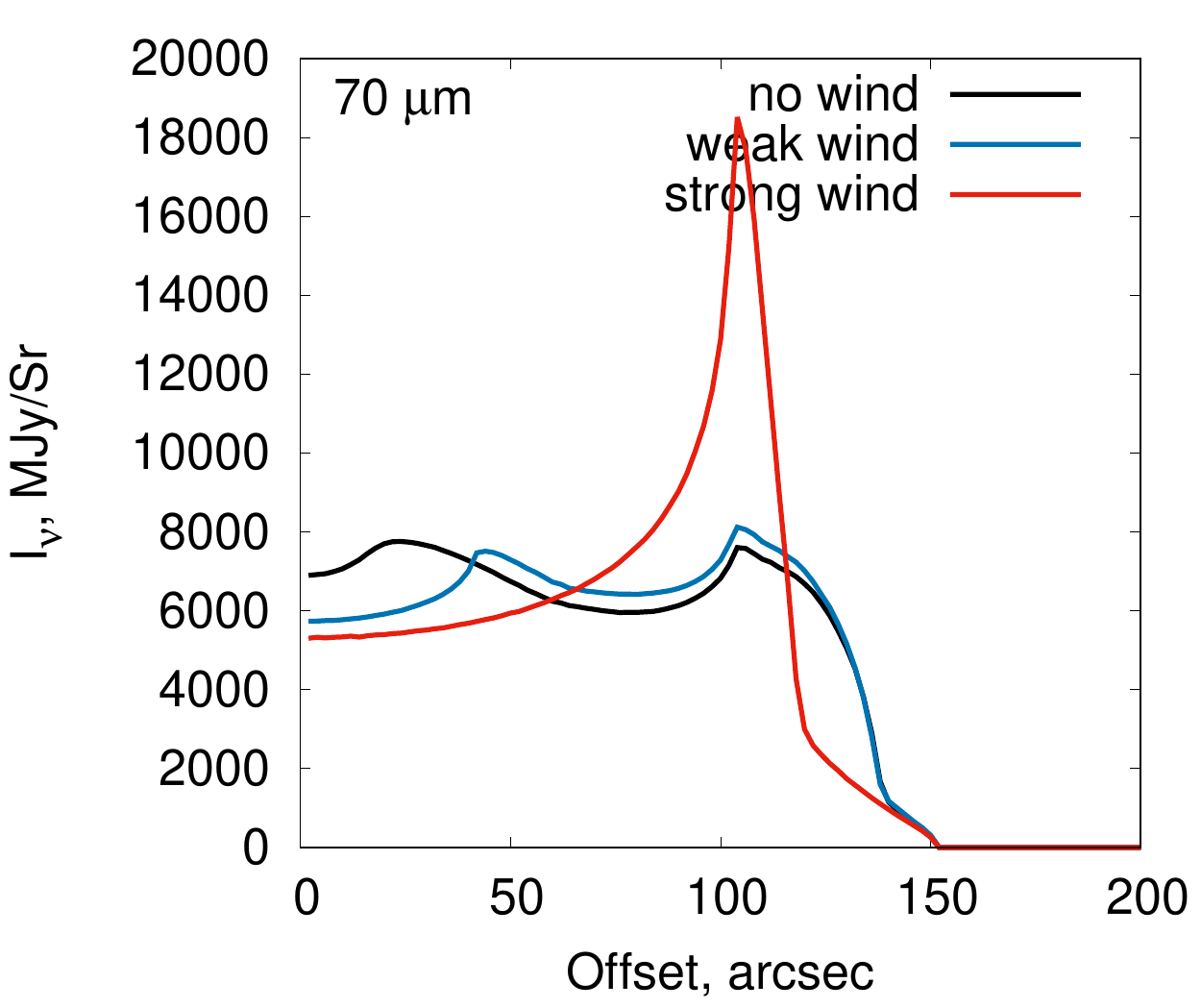}
    \includegraphics[width=0.45\columnwidth]{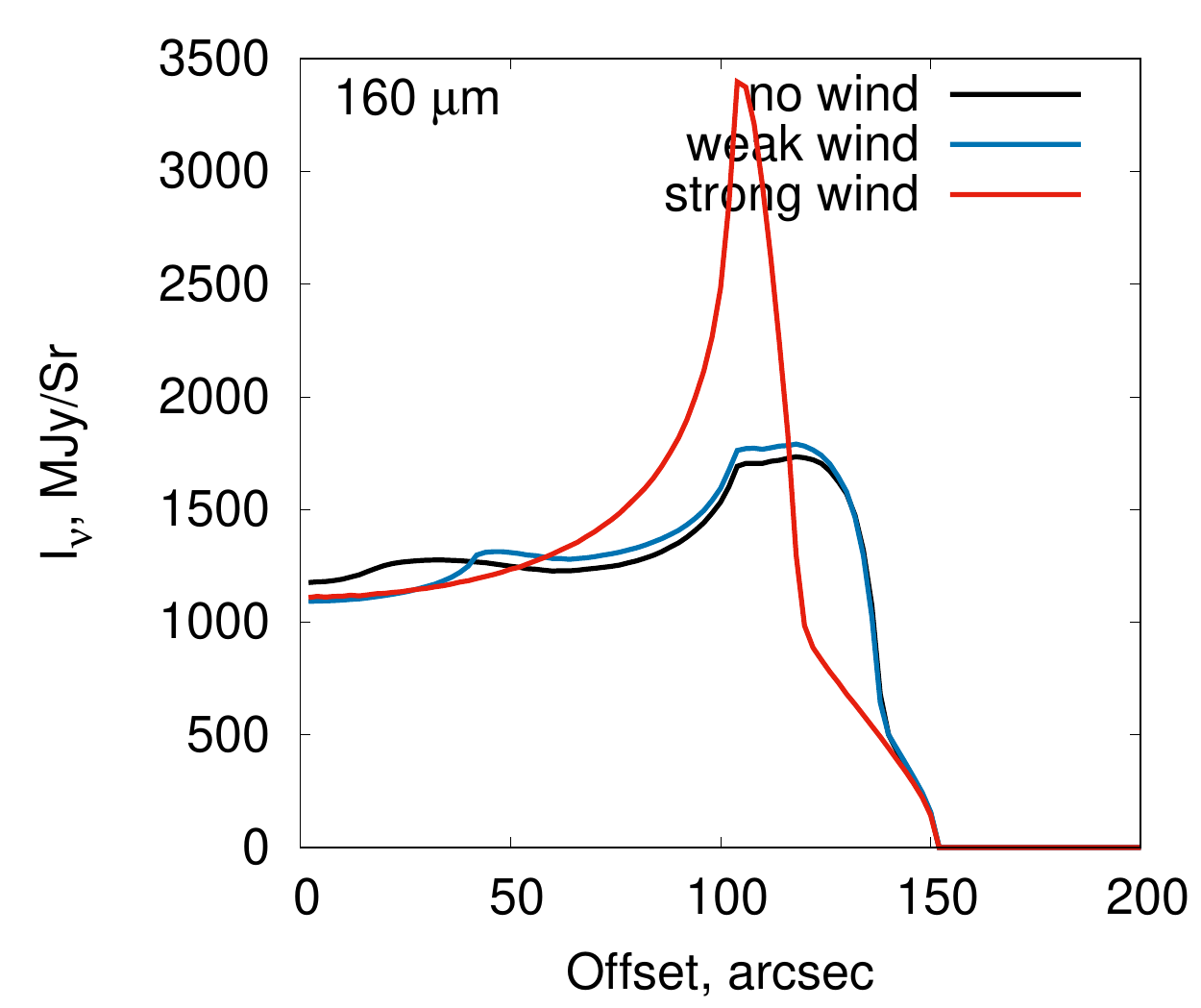}\\
    \includegraphics[width=0.45\columnwidth]{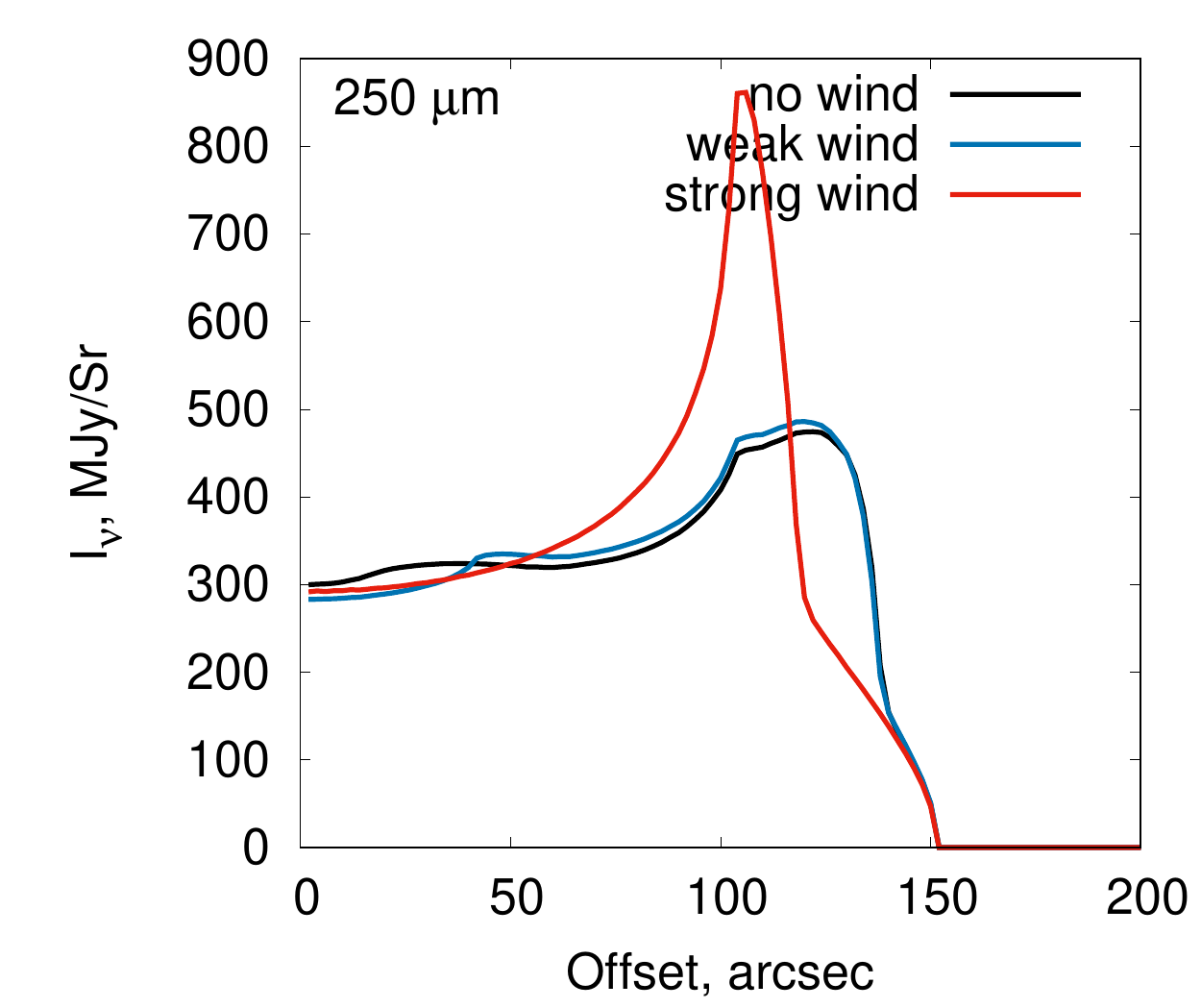}
    \includegraphics[width=0.45\columnwidth]{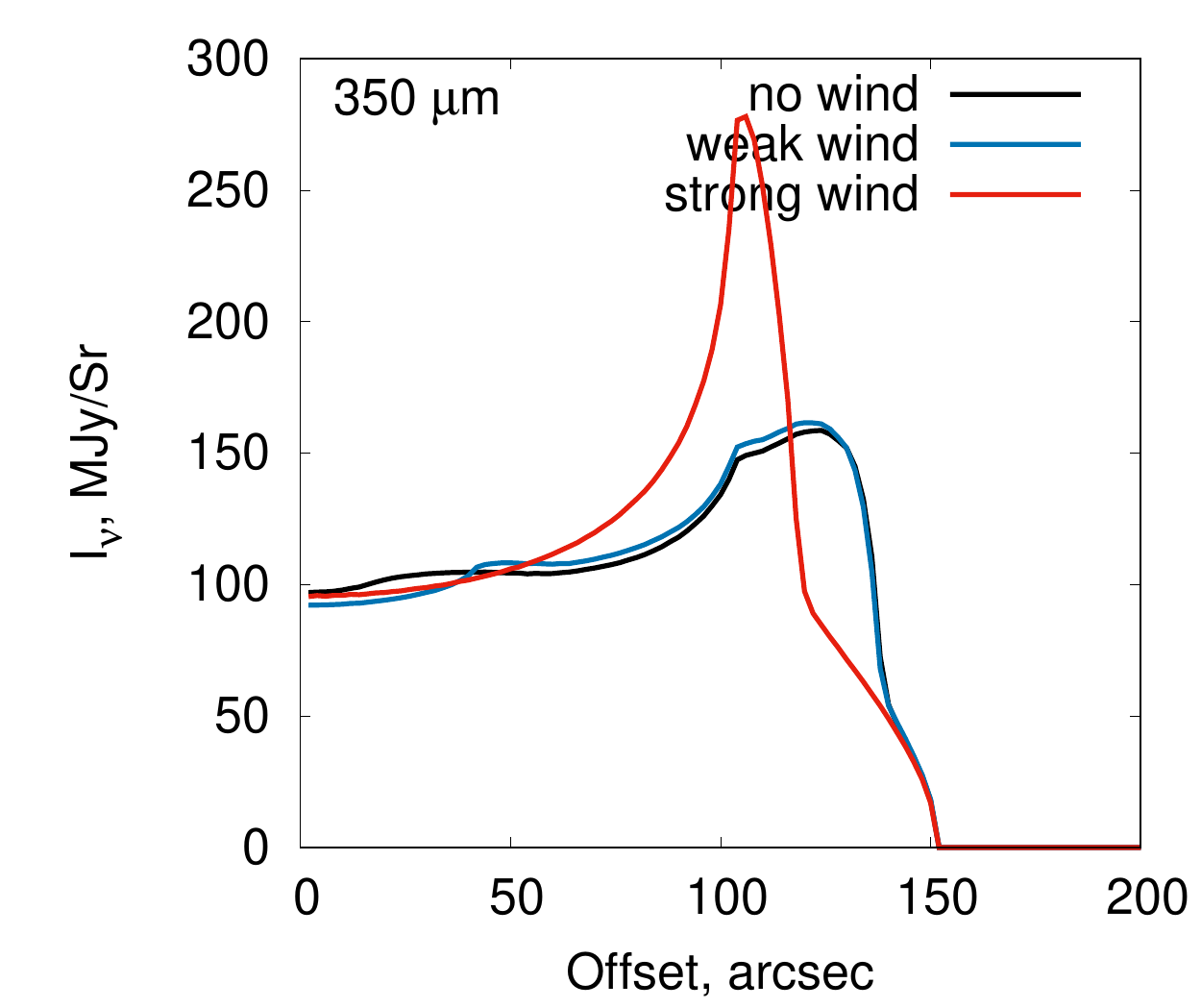}
    \caption{Radial intensity profiles at 8~$\mu$m, 24~$\mu$m, 70~$\mu$m, 160~$\mu$m, 250~$\mu$m and 350~$\mu$m calculated for the models with and without stellar wind.}
    \label{figdust}
    \end{center}
\end{figure}

Comparing simulated distributions of infrared emission in three considered models, we find the most prominent differences at 8 and 24~$\mu$m, see Fig.~\ref{figdust}. There is a bright peak of the 8~$\mu$m emission near the centre of \hii{} region in the model without wind, related to small radial dust drift under the effect of radiation pressure. This effect alone does not clear the \hii region, therefore simulations with no wind can not explain typical observed distributions of the 8~$\mu$m even qualitatively. The weak wind model also produces a peak at 8~$\mu$m inside of the bubble, while the strong wind model evacuates all PAHs from the \hii{} region and carries out only the ring-like 8~$\mu$m emission.  

On the contrary, the simulated 24~$\mu$m distribution is qualitatively consistent with the observed morphology in the models without and with weak wind. They both produce the bright inner ring around the ionising star and the less bright outer ring related with the dense shovelled envelope. The strong wind model produces a nonobservable one ring-like structure at the dense envelope because the bubble is empty.

All three considered models produce ring-like emission at 70~$\mu$m and longer wavelengths with approximately flat intensity distribution within \hii{} region. Strong wind makes the shovelled neutral envelope thin and dense, therefore the far-infrared emission is two times brighter than in the other two models. We conclude that the far-infrared dust emission is not sensitive to the stellar wind parameters.

\section{Conclusion}

Different distributions of the 8 and 24~$\mu$m emission around young massive stars can be attributed to the specific parameters of stellar wind. There is a bright 8~$\mu$m emission from \hii{} regions in the models without or with the weak stellar wind. Therefore, PAHs should be removed from the \hii{} regions by some other way to produce the commonly observed 8~$\mu$m ring. Ultraviolet photons can destroy PAHs, and we consider this way as one of the possibilities to reduce brightness at 8~$\mu$m inside of the \hii{} regions. For example, \cite{Pavyar_etal13} proposed phenomenological model of the PAH destruction by stellar UV photons and found that this mechanism can be effective to create the 8~$\mu$m rings. However, subsequent detailed modelling by \cite{Murga_etal22} confirmed the importance of stellar UV photons only for small PAHs with sizes less than $\approx 6-7$~\AA. Extreme UV or ever soft X-ray photons could be effective to destroy larger PAHs, therefore we consider this direction as promising continue of our numerical study. Central peaks at 24~$\mu$m can be produced in the models without and with the weak wind. All dust types are removed from \hii{} regions in the strong wind model. Far-IR dust emission is not sensitive to the stellar wind parameters.

\section{Acknowledgements} We are thankful to S.~Yu. Parfenov and J. Mackey for fruitful discussions. We also thank referee for valuable comments and suggestions. The study is supported by Russian Science Foundation (project 21-12-00373).

\begin{discussion}

\discuss{Mackey}{It seems like in your models the dust and the gas are very well coupled. Maybe this is because of the quite high density. Did you look at some cases where there is quite significant drift between the gas and the dust? Have you explored that?}

\discuss{Kirsanova}{Yes, they are coupled, but the dust is not frozen to the gas. We obtain empty \hii{} regions in the models with stellar wind. If we don't include the wind, we see significant differences. We calculate models with quite high density in order to obtain quantitative agreement between our synthetic observations in the mid and far-IR and the observed values.} 

\end{discussion}

\end{document}